\pdfoutput=1
\documentclass[prd,onecolumn,notitlepage,eqsecnum,nofootinbib,floatfix,superscriptaddress]{revtex4-1}
\usepackage{amssymb}
\usepackage{amsmath}
\usepackage{amsfonts} 
\usepackage{bm}
\usepackage{color} 
\usepackage{graphicx}
\usepackage[bookmarks=false]{hyperref} %make sure it comes last of your loaded packages
\hypersetup{pdfstartview=FitH,pdfhighlight=/O,colorlinks=true}

\allowdisplaybreaks

\newtheorem{theorem}{Result}
\newtheorem{conjecture}{Conjecture}
\usepackage{epsfig} 
\usepackage{epstopdf}
\usepackage{latexsym}
\usepackage{psfrag}   
\usepackage{subfigure}  
\usepackage{booktabs}  
\usepackage{braket}
\usepackage{mathtools}
\usepackage{textcomp}
\usepackage{ifthen}
\usepackage{tensor}
\usepackage{comment}
\usepackage{protosem} 
\usepackage{wasysym}
\usepackage[inner=2.9cm,outer=2cm]{geometry}
\usepackage[toc]{appendix}
\usepackage{color,soul} 
\usepackage{datetime}
\definecolor{darkblue}{rgb}{0.2, 0, 0.8}
\definecolor{darkgreen}{rgb}{0.2, 0.71, 0}
\definecolor{awesome}{rgb}{1.0, 0.13, 0.32}
\definecolor{cadmiumred}{rgb}{0.89, 0.0, 0.13}
\definecolor{dukeblue}{rgb}{0.0, 0.0, 0.61}

\numberwithin{equation}{section}

\newcommand{\req}[1]{(\ref{#1})} %{Eq.\thinspace(\ref{#1})}  

\newcommand{\bea}{\begin{eqnarray}}
\newcommand{\eea}{\end{eqnarray}}
\newcommand{\ba}{\begin{eqnarray}}
\newcommand{\ea}{\end{eqnarray}}

\newcommand{\beq}{\begin{equation}}
\newcommand{\eeq}{\end{equation} }
\newcommand{\beqa}{\begin{eqnarray}}
\newcommand{\eeqa}{\end{eqnarray}}
\newcommand{\beqar}{\begin{eqnarray*}}
\newcommand{\eeqar}{\end{eqnarray*}}

\newcommand{\ssc}{\scriptscriptstyle}
\newcommand{\eg}{{\it e.g.,}\ }
\newcommand{\ie}{{\it i.e.,}\ }

\renewcommand{\href}[2]{#2}

\begin{document}
	\title{On black holes in higher-derivative gravities}
	\author{Pablo Bueno}
	\email{pablo@itf.fys.kuleuven.be}
	\affiliation{Instituut voor Theoretische Fysica, 
	KU Leuven, Celestijnenlaan 200D, B-3001 Leuven, Belgium}
	\author{Pablo A. Cano}
	\email{pablo.cano@uam.es}
	\affiliation{Instituto de F\'isica Te\'orica UAM/CSIC,
		C/ Nicol\'as Cabrera, 13-15, C.U. Cantoblanco, 28049 Madrid, Spain}

\begin{abstract}
	We establish various general results concerning static and spherically symmetric black hole solutions of general higher-derivative extensions of Einstein gravity. We prove that the only theories susceptible of admitting solutions with $g_{tt}g_{rr}=-1$ and representing the exterior field of a spherically symmetric distribution of mass are those that only propagate a massless and traceless graviton on the vacuum. Then, we provide a simple --- and computationally powerful --- sufficient condition for a theory to admit solutions of that kind, as well as a systematic way for constructing them for a given theory. We conjecture (and provide strong evidence) that all black holes constructed according to our criteria are completely determined by their mass (non-hairy), and such that their thermodynamic properties can be obtained by solving a system of algebraic equations without free parameters. Our results can be straightforwardly extended to planar and hyperbolic horizons. We illustrate this by obtaining new planar asymptotically $AdS_5$ black hole solutions of the recently constructed \emph{Generalized quasitopological gravity} [arXiv:1703.01631], which belongs to the class of theories selected by our results.

\end{abstract}

\maketitle
%\tableofcontents

%%%%%%%%%%%%%%%%%%%%%%%%%%%%%%
\section{Introduction \& summary of results} 
\label{sec:Introduction} 
%%%%%%%%%%%%%%%%%%%%%%%%%%%%%%

Higher-derivative gravities play an important role in various areas of (high-energy) physics. These include, among others: cosmology, black hole physics, holography, supergravity and, more broadly, string theory. Note, in particular, that the Einstein-Hilbert term is expected to be just the first of an infinite set of higher-derivative contributions appearing in the (stringy) gravitational effective action --- see \eg \cite{Gross:1986mw,Green:2003an,Prue}.

Recently, higher-derivative gravities have attracted a lot of interest in the context of AdS/CFT \cite{Maldacena,Witten,Gubser}. In that framework, these theories can be used to construct holographic CFTs whose properties are distinct from those corresponding to Einstein gravity duals \cite{Brigante:2007nu,deBoer:2009pn,Camanho:2009vw,Buchel:2009tt,Cai:2009zv,Camanho:2009hu,Buchel:2009sk,Myers:2010jv,Quasi}. Such theories have proven to be particularly useful in some cases, to the extent of playing a crucial role in the discovery of new universal properties valid for general CFTs \cite{Myers:2010xs,Myers:2010tj,Bueno1,Bueno2,Mezei:2014zla}.

From a more generic perspective, the study of higher-derivative gravities allows for a better understanding of Einstein gravity and its properties, as it provides a framework for testing what features of the theory are special, and which ones persist when additional terms are considered in the action.

In this paper we will obtain several results regarding the black hole solutions of higher-derivative theories of the form\footnote{Throughout the paper we often refer to this class of theories as \emph{higher-derivative} gravities. Sometimes we use \emph{higher-order gravities} to denote the subclass for which $\mathcal{L}$ does not depend on covariant derivatives of the Riemann tensor. }
\begin{equation}\label{higherd}
S=\int d^Dx \sqrt{|g|}\, \mathcal{L}(g^{ab},R_{abcd},\nabla_{e}R_{abcd},\ldots)\, .
\end{equation}
Throughout the paper, we will be mostly interested in extensions of Einstein gravity, \ie we will assume that the above action reduces to the Einstein-Hilbert one when all higher-derivative couplings are set to zero. In that case, the above Lagrangian can be written in the form
\begin{equation}\label{higherd}
\mathcal{L}(g^{ab},R_{abcd},\nabla_{e}R_{abcd},\ldots)=\frac{1}{16\pi G}\left[-2\Lambda_0+R+\text{higher-derivative terms}\right]\, ,
\end{equation}
where the higher-derivative terms are assumed to be arbitrary linear combinations of polynomials of the Riemann tensor and its covariant derivatives. 

In particular, our interest will be on static and spherically symmetric solutions of the form\footnote{Metrics of this form have Ricci tensors with vanishing radial null-null components, see \eg \cite{Salgado:2003ub,Jacobson:2007tj}.}
\begin{equation}
 ds^2_f=-f( r )dt^2+\frac{dr^2}{f( r )}+r^2 d\Omega_{(D-2)}^2\, .
 \label{Fmetric}
 \end{equation}
These will be generalizations of the well-known Schwarzschild-Tangherlini-(A)dS solution (which solves the field equations of \req{higherd} when all the higher-derivative couplings are set to zero),\footnote{We will sometimes refer to \req{Fmetric} with \req{SAdS} as `Schwarzschild-(A)dS' solution, by which we will be referring to the three possible asymptotic behaviors: Anti-de Sitter ($\Lambda_0<0$), de Sitter ($\Lambda_0>0$) or flat ($\Lambda_0=0$).} for which
\begin{equation}\label{SAdS}
f(r)=1-\frac{16\pi GM}{(D-2)\Omega_{(D-2)}r^{D-3}}-\frac{2\Lambda_0r^2}{(D-1)(D-2)}\, , \quad \text{where} \quad \Omega_{(D-2)}=\frac{2 \pi^{\frac{D-1}{2}}}{\Gamma[\frac{D-1}{2}]}
\end{equation}
is the area of the $(D-2)$-dimensional unit sphere
and $M$ is the ADM mass \cite{Arnowitt:1960es,Arnowitt:1960zzc,Arnowitt:1961zz}. Although our focus will be on spherically symmetric solutions, many of our results will be easily generalizable to planar and hyperbolic horizons, as we will explicitly illustrate --- see below.

On general grounds, finding static and spherically symmetric black hole solutions for $D$-dimensional theories of the form \req{higherd} is a challenging task. In particular, if the Lagrangian contains up to $n$ derivatives of the Riemann tensor, the equations of motion generally involve $2n+4$ derivatives of the metric which, in the case of a general static and spherically symmetric ansatz --- see \req{Nf} below --- usually translate into a system of coupled differential equations of such order --- see \eg \cite{Lu:2015psa,Lu:2015cqa}. As a matter of fact, some simple analytic solutions of the form \req{Fmetric}, \ie
characterized by the condition $g_{tt}g_{rr}=-1$, have been in fact constructed for certain higher-derivative theories. However, as we pointed out in \cite{PabloPablo2} these solutions fall, very often, within one of the following three categories:
\begin{itemize}
\item[i)] They are the `same' solution as in Einstein gravity, \ie they correspond to embeddings of Einstein gravity solutions in some higher-derivative theory --- see \eg \cite{delaCruzDombriz:2009et} for the prototypical example of Schwarzschild-(A)dS in $f(R)$ gravity.
\item[ii)] They are solutions to pure higher-derivative gravities, so that the action does not include the Einstein-Hilbert term --- and hence they lack an Einstein gravity limit. For instance, pure Weyl-squared gravity, $\mathcal{L}=\alpha C_{abcd}C^{abcd}$, in $D=4$ allows for solutions of the form \req{Fmetric} \cite{Riegert:1984zz,Klemm:1998kf}, while $\mathcal{L}=-2\Lambda_0+R+\alpha C_{abcd}C^{abcd}$ does not \cite{Lu:2012xu}. Examples of this kind involving Weyl-cubed terms in $D=6$ can be found \eg in \cite{Oliva:2010zd}. Similar comments apply to pure Lovelock gravity solutions, like those constructed \eg in \cite{Banados:1993ur,Cai:2006pq}.
\item[iii)] They involve the fine-tuning of some of the higher-derivative couplings --- and hence, again, they lack an Einstein gravity limit. A simple example corresponds to perfect-square (or other powers greater than $2$) actions, \eg $\mathcal{L}=-(R-4\Lambda_0)^2/(8\Lambda_0)$, which of course admits solutions of the form $R=4\Lambda_0$.  Examples belonging to this class have been constructed, \eg in \cite{Cai:2009ac,Love}.
\end{itemize}
We will not consider cases ii) and iii) because of their lack of Einstein gravity limit. With regards to case i), we will show that these solutions are in fact ``unnatural'', in the sense that they cannot correspond to the exterior gravitational field of generic spherically symmetric distributions of mass --- see next subsection and section \ref{unnatural} for details.

On the bright side, genuine (analytic or semianalytic) single-function extensions of \req{SAdS} have been constructed in $D\geq 5$ for Lovelock gravities \cite{Wheeler:1985nh,Wheeler:1985qd,Boulware:1985wk,Cai:2001dz,Dehghani:2009zzb,deBoer:2009gx}, Quasitopological gravity \cite{Quasi2,Quasi} and its quartic \cite{Dehghani:2011vu} and quintic \cite{Cisterna:2017umf} generalizations. The $D=4$ case turns out to be a considerably harder nut to crack, given that all the Lovelock and Quasitopological densities (except for the Einstein-Hilbert terms) are either topological or trivial in that case. Recently, the first four-dimensional generalizations to \req{SAdS} of the form \req{Fmetric} have been constructed in \cite{Hennigar:2016gkm,PabloPablo2} for Einsteinian cubic gravity \cite{PabloPablo}, a theory which was  originally identified because of the special properties of its linearized spectrum in general dimensions --- see footnote \ref{ECGf}.
Even more recently, additional solutions have been obtained after the construction of \emph{Generalized quasitopological gravity} \cite{Hennigar:2017ego} which, in a way, includes both $D=4$ Einsteinian cubic gravity and Quasitopological gravity as particular limits. The main goal of this paper is to identify and characterize some of the properties which make all these theories special as well as to provide explanations for some previously conjectured results. We summarize our findings in the following subsection.

\subsection{Main results}\label{main}
Most of the material presented in sections \ref{bhl} and \ref{bho} can be encapsulated in two main results 
%\footnote{Note that we have refrained form using the word ``Theorem'' to refer to our ``Results''. While we think our proofs are completely robust, we have considered that some additional level of rigor could perhaps be added  }
 (which we prove) and two conjectures (in favor of which we provide strong evidence). In order to formulate them, let us start with a couple of definitions. First, let $L_{N,f}$ be the effective Lagrangian resulting from the evaluation of $\sqrt{|g|}\mathcal{L}$ in
the general static and spherically symmetric ansatz
\begin{equation}\label{Nf}
ds^2_{N,f}=-N^2( r )f( r )dt^2+\frac{dr^2}{f( r )}+r^2 d\Omega_{(D-2)}^2\, .
\end{equation}
More precisely, we define
\begin{equation}\label{LNf}
L_{N,f}(r,f(r), N( r ),f'(r), N'( r ), \ldots)\equiv N( r )r^{D-2}\mathcal{L}\big|_{g^{ab}=g^{ab}_{N,f}}\, .
\end{equation}
%where $\Omega_{(D-2)}$ is the area of $S^{(D-2)}$.
Analogously, we will denote by $L_f$ the expression resulting from setting $N=1$ in $L_{N,f}$, which of course corresponds to the effective Lagrangian for $f(r)$ resulting from the evaluation of $\sqrt{|g|}\mathcal{L}$ in the single-function ansatz (\ref{Fmetric}). With these definitions at hand, we are ready to enumerate our results:

\begin{theorem}\label{ti1}
 Given a $ \mathcal{L}(g^{ab},R_{abcd})$ gravity of the form \req{higherd} such that: 1) It allows for static and spherically symmetric solutions characterized by a single function (\ref{Fmetric}); 2) These solutions represent the exterior gravitational field of a spherically symmetric mass distribution, then:
 \begin{enumerate}
\item The theory only propagates a traceless and massless graviton on the vacuum.\footnote{We assume that the vacuum is a maximally symmetric space.}
 \end{enumerate}
 \end{theorem}
%\\
%This theorem is telling us that, in order to obtain theories whose solutions are characterized by a single function and that satisfy the reasonable requirement of coinciding with an exterior solution of a mass distribution, we must search among the Einstein-like theories.\\

\begin{theorem} \label{theo}Given a higher-derivative gravity Lagrangian $\mathcal{L}(g^{ab},R_{abcd},\nabla_{e}R_{abcd},\ldots)$ of the form \req{higherd} involving terms with up to $n$ covariant derivatives of the Riemann tensor. If the Euler-Lagrange equation of $L_f$ vanishes identically, \ie if 
\begin{equation}\label{EELL}
\frac{\delta L_f}{\delta f}\equiv\frac{\partial L_f}{\partial f}-\frac{d}{dr}\frac{\partial L_f}{\partial f'}+\frac{d^2}{dr^2}\frac{\partial L_f}{\partial f''}-\ldots= 0\quad \forall \, f(r)\, ,
\end{equation}
 then: 
 \begin{enumerate}
 	\item{The theory allows for solutions of the form (\ref{Fmetric}), where $f(r)$ satisfies a differential equation of order $\le 2n+2$.}
 	\item{The former solutions coincide with the exterior gravitational field of a spherically symmetric body, and they are fully characterized by the total mass $M$.}
 	\item{At least in the $ \mathcal{L}(g^{ab},R_{abcd})$ case, the theory only propagates a traceless and massless graviton on the vacuum.}
 \end{enumerate}
\end{theorem}

  \begin{conjecture}\label{conj} For all theories fulfilling the hypothesis of Result \ref{theo}:
  	
  	\begin{enumerate}
  		\item{(No-hair) There is a unique black hole solution of the form (\ref{Fmetric}) solely characterized by its mass $M$.}
  		\item{The thermodynamic properties (\eg $T$, $\mathsf{S}$) of this black hole can be determined by solving a system of algebraic equations without free parameters.}
  	\end{enumerate}
  \end{conjecture}

 \begin{conjecture}\label{conji2} Given a $\mathcal{L}(g^{ab},R_{abcd})$ theory fulfilling the hypothesis of Result \ref{theo}, if $L_{N,f}$ can be written as
 	\begin{equation}\label{siss}
 	L_{N,f}=N L_f+ N' F_1+N'' F_2\, ,
 	\end{equation} 
 	where $F_{1,2}$ are functions of $f(r)$ and its derivatives, then:
 	\begin{enumerate}
 		\item{The equation determining $f(r)$ is algebraic.}
 	\end{enumerate}
 \end{conjecture}
Some comments are in order. 

\begin{itemize}

	\item[-] Result \ref{ti1}, which we prove in section \ref{bhl1}, explains the previously noticed \cite{Quasi,PabloPablo,Hennigar:2016gkm,PabloPablo2,Aspects,Cisterna:2017umf,Hennigar:2017ego} (but so far unexplained) fact, that certain higher-order gravities admitting simple black hole solutions (in particular, of the form \req{Fmetric}), have the interesting property of sharing the linearized spectrum of Einstein gravity. A crucial hypothesis in our result is the assumption that the corresponding solutions describe the exterior field of spherically symmetric mass distributions. The apparent contradiction of Result \ref{ti1} with the fact that certain theories which propagate extra modes at the linearized level do admit black hole solutions of the form \ref{bhl} is not such. The reason is that, in those cases, the corresponding solutions cannot correspond to the exterior gravitational field of generic spherical distributions, which makes them somewhat ``unnatural''. We illustrate this very explicitly for the well-known Schwarzschild-(A)dS solution in $f(R)$ gravity in section \ref{unnatural}. 

 	\item[-] While Result \ref{ti1} provides a necessary condition for a theory to allow for solutions characterized by a single function and representing the exterior field of a spherical body, Result \ref{theo}, which we prove in section \ref{bho1}, yields a sufficient condition for a theory to satisfy this requirement. Interestingly, the proof of the first item of Result \ref{theo}, provides a very efficient method for identifying higher-derivative gravities allowing for solutions of the form \req{Fmetric} as well as for obtaining the differential equation determining $f(r)$ in each case. We present this method  in the form of a simple recipe in section \ref{recipe}. 
 	
	\item[-] Note that the reason for considering the subclass of theories $\mathcal{L}(g^{ab},R_{abcd})$ (which do not include terms involving covariant derivatives of the Riemann tensor), in Result \ref{ti1} and in epigraph 3 of Result \ref{theo} is that the spectrum and Newtonian limit of these theories has been exhaustively classified in \cite{Aspects}, which we use to prove our results. We are not aware of an analogous general classification in the general higher-derivative case. However, in light of some related recent works \cite{Modesto:2014eta,Giacchini:2016xns}, we strongly believe those results apply in the general case as well.
	
 	\item[-] Observe that Result \ref{theo} does not really make reference to whether the solutions described by \req{Fmetric} in each case can correspond, in particular, to black holes. However, the great amount of evidence accumulated so far \cite{Wheeler:1985nh,Wheeler:1985qd,Boulware:1985wk,Cai:2001dz,Dehghani:2009zzb,Quasi,Quasi2,deBoer:2009gx,Dehghani:2011vu,PabloPablo,Hennigar:2016gkm,PabloPablo2,Cisterna:2017umf,Hennigar:2017ego}, along with our results, provide strong support for the validity of Conjecture \ref{conj}.

	\item[-]  For a $\mathcal{L}(g^{ab},R_{abcd})$ theory satisfying the hypothesis of Result \ref{ti1}, the order of the differential equation determining $f(r)$ is usually $2$ --- see \eg \cite{Hennigar:2016gkm,PabloPablo2,Hennigar:2017ego}. However, in some well-known cases
	\cite{Wheeler:1985nh,Wheeler:1985qd,Boulware:1985wk,Cai:2001dz,Dehghani:2009zzb,deBoer:2009gx,Quasi,Quasi2,Dehghani:2011vu,Cisterna:2017umf}, the equation is algebraic instead, which represents a considerable simplification. Conjecture \ref{conji2}, which we motivate in sections \ref{quadratic} and \ref{cubic}, provides a straightforward guiding principle for identifying such class of theories from a given larger set.

 	\item[-] Finally, note that, even though we focus on the spherically symmetric case, the solutions which can be constructed using our method can be straightforwardly generalized to the hyperbolic and flat transverse geometry cases. In particular, the recipe explained in section \ref{recipe} can equally be applied to those cases. We illustrate this in section \ref{planar}, where we obtain new five-dimensional asymptotically AdS planar black hole solutions of the recently constructed Generalized quasitopological gravity \cite{Hennigar:2017ego} --- we also comment on this theory in section \ref{cubic}. This also serves to further support Conjecture \ref{conj} and is of course motivated by holography, where planar black hole solutions to various higher-order gravities have proven to be remarkably useful for different purposes --- see	\eg \cite{Brigante:2007nu,deBoer:2009gx,Buchel:2009sk,Myers:2010jv,Myers:2010tj}.

\end{itemize}

 \section{Black holes and linearized gravity}\label{bhl}
  It has been previously observed that certain higher-order gravities admitting simple black hole solutions possess particularly simple linearized spectra. This was emphasized by Myers and Robinson in \cite{Quasi}, where they observed that Quasitopological gravity \cite{Quasi2,Quasi} satisfies the following two unusual properties: first, it admits black holes characterized by a single function $f(r)$, \ie solutions of the form \req{Fmetric}; and second, its linearized spectrum coincides with the Einstein gravity one, namely, the only dynamical mode propagated by the metric perturbation in a maximally symmetric spacetime is a transverse and traceless graviton. These two, apparently unrelated, properties are also known to hold for general Lovelock theories \cite{Lovelock1,Lovelock2,Wheeler:1985nh,Wheeler:1985qd,Boulware:1985wk,Cai:2001dz,Dehghani:2009zzb,deBoer:2009gx,Love}, for Einsteinian cubic gravity in four dimensions \cite{PabloPablo,Hennigar:2016gkm,PabloPablo2}, for Quartic \cite{Dehghani:2011vu} and Quintic quasitopological gravity \cite{Cisterna:2017umf}, and for the recently constructed Generalized quasitopological gravity \cite{Hennigar:2017ego}. Hence, it is natural to wonder how generally this connection between the linearized regime and the genuinely non-linear one holds for general higher-order gravities. 
  
  It is important to note that not all higher-order theories which share the linearized spectrum of Einstein gravity admit single-function black hole solutions. Examples of such theories include 
  Einsteinian cubic gravity in $D\geq 5$ \cite{Hennigar:2016gkm,PabloPablo2} and certain $f($Lovelock$)$ theories \cite{Love,Aspects,Karasu:2016ifk}. Furthermore, there are theories which propagate extra modes at the linearized level and yet they posses solutions of the form \req{Fmetric}. This is the case, \eg of $f(R)$ gravity \cite{delaCruzDombriz:2009et}. Naturally, these observations clearly show that the connection between both properties cannot be a double implication.
\subsection{Single-function black holes and linearized spectrum}\label{bhl1}

 In this section we will prove that, in fact, given a general $ \mathcal{L}(g^{ab},R_{abcd})$ theory, if the exterior gravitational field of a spherically symmetric body is given by a metric of the form \req{Fmetric},\footnote{These solutions will often correspond to black holes, but notice that Result \ref{ti1} holds without the extra assumption that \req{Fmetric} actually corresponds to a black hole metric. Of course, it also holds in that particular case.}
then the theory only propagates a massless graviton at the linearized level. In other words, only theories sharing the linearized spectrum of Einstein gravity are susceptible of admitting 
single-function (black-hole) solutions corresponding to the exterior field of a spherically symmetric body. 

 In order to prove this statement, let us consider the linearization of a general $ \mathcal{L}(g^{ab},R_{abcd})$ theory on a maximally symmetric background. In \cite{Aspects}, we proved that the Newtonian metric of a point-particle of mass $M$ in any such theory in $D$-dimensions is given by
 \begin{equation}
 	ds_N^2=-(1+2U(\rho))dt^2+(1-2V(\rho))(d\rho^2+\rho^2 d\Omega_{(D-2)}^2)\,,
 	\label{Newtonian}
 \end{equation}
 where $U(\rho)$ is the generalized Newton potential and $V(\rho)=\gamma(\rho) U(\rho)$ where $\gamma(\rho)$ is one of the so-called parametrized Post-Newtonian parameters. In four dimensions, these functions are explicitly given by\footnote{In general dimensions, the equivalent expressions read \cite{Aspects}
  \begin{align}\label{uu}
  U( \rho )&=-\mu(D) \frac{G_{\rm{eff}} M}{\rho^{D-3}}\left[1+\nu(D)\rho^{\frac{D-3}{2}}\left[-m_g^{\frac{D-3}{2}}K_{\frac{D-3}{2}}(m_g \rho)+\frac{m_s^{\frac{D-3}{2}}}{(D-2)^2}K_{\frac{D-3}{2}}(m_s \rho) \right]\right]\, , \\ 
  \label{uv}
  \gamma( \rho )&=\frac{1-\frac{2}{(D-1)\Gamma({\frac{D-3}{2}})}\Big[(D-2)\left(\frac{m_g \rho}{2}\right)^{\frac{D-3}{2}}K_{\frac{D-3}{2}}(m_g \rho)
  	+\left(\frac{m_s \rho}{2}\right)^{\frac{D-3}{2}}K_{\frac{D-3}{2}}(m_s \rho)\Big]}{D-3-\frac{2}{(D-1)\Gamma({\frac{D-3}{2}})}\Big[(D-2)^2\left(\frac{m_g \rho}{2}\right)^{\frac{D-3}{2}}K_{\frac{D-3}{2}}(m_g \rho)
  	-\left(\frac{m_s\rho}{2}\right)^{\frac{D-3}{2}}K_{\frac{D-3}{2}}(m_s \rho)\Big]}\, ,
  \end{align}
  with
  $ \mu(D) \equiv  8 \pi/( (D-2) \Omega_{D-2})$, $\nu(D) \equiv(D-2)^2/(\Gamma\left[{\frac{D+1}{2}}\right]2^{\frac{D-1}{2}})$. Also, $K_{\ell}(z)$ is the modified Bessel function of the second kind.  }
 
 \begin{equation}
 	U ( \rho )=-\frac{G_{\rm{eff}}M}{\rho}\left[1-\frac{4}{3}e^{-m_g \rho}+\frac{1}{3}e^{-m_s \rho}\right],\quad  V ( \rho)=-\frac{G_{\rm{eff}}M}{\rho}\left[1-\frac{2}{3}e^{-m_g \rho}-\frac{1}{3}e^{-m_s \rho}\right].
 	\label{potentials}
 \end{equation}
 In these expressions, $m_g$ and $m_s$ are, respectively, the masses of the additional spin-2 and spin-0 modes propagated by the metric perturbation for a generic $ \mathcal{L}(g^{ab},R_{abcd})$ theory. These can be easily computed for a given theory using the method developed in \cite{PabloPablo,Aspects}.\footnote{See also \cite{Tekin1,Tekin2,Tekin4} for previous works on the linearization of higher-order gravity theories on maximally symmetric backgrounds.}

The Newtonian metric \req{Newtonian} is written in isotropic coordinates. In order to express it in Schwarzschild-like coordinates, we perform the change of variable $r^2=\rho^2(1-2V( \rho))$.  We stay at linear order in $V$ and $U$, which suffices for our purposes. Note that $V(\rho)=V( r )+\mathcal{O}(V^2)$ and the same is true for $U$. Then, we obtain
 \begin{equation}\label{tte}
 	ds_N^2=-(1+2 U( r ))dt^2+(1+2r V'( r ))dr^2+r^2 d\Omega_{(D-2)}^2\,.
 \end{equation}
 Now we make use of our hypothesis and  assume that this is indeed the linearized limit of a full non-linear solution of the form (\ref{Fmetric}). This means that $g_{tt}g_{rr}=-1$, which, when  applied to \req{tte}, imposes the following condition on $U$ and $V$,
 \begin{equation}
 	U( r )+ r V'( r )=0\, .
 	\label{condition1}
 \end{equation}
 From the expressions in (\ref{potentials}), it follows that this condition holds only when $m_g^2=m_s^2=+\infty$. In other words, the single-function condition on the non-linear solution
implies the absence of the massive graviton and the scalar field, so that the only mode which is propagated on the vacuum is the massless graviton. The argument extends straightforwardly to general dimensions using \req{uu} and \req{uv}.
%In general dimensions $D>4$ the specific expressions for $U$ and $V$ are different, but the conclusion is the same; (\ref{condition1}) only holds whenever the scalar and the massive graviton are removed.
This simple argument shows that if a higher-order gravity allows for a solution of the form (\ref{Fmetric}) representing the exterior field of a spherical body, the theory only propagates a massless graviton on the vacuum. 

Note that, strictly speaking, the Newtonian metric (\ref{Newtonian}) only applies in the asymptotically flat case. In that situation, only terms up to quadratic order in curvature contribute to the masses $m_g$ and $m_s$, so our argument does not immediately go through beyond the quadratic level. In order to extend it to terms of arbitrary order in curvature, we must generalize this argument to include asymptotically (A)dS solutions, since in that case all terms do contribute to the masses of the modes. For a background of curvature $\Lambda$, the Newtonian solution (\ref{Newtonian}) is also a good approximation as long as $\rho<<|\Lambda|^{-1/2}$. Hence, the same analysis can be applied and the same conclusions are reached, namely, $m_g^2=m_s^2=+\infty$, where now the masses contain information about terms at every order in curvature. Alternatively, one can compute explicitly the Newtonian potential in the (A)dS case and repeat the analysis, but the conclusions remain unchanged.

As we have seen, the results in this section rely on the analysis performed in \cite{PabloPablo,Aspects} for the Newtonian metric of a general $ \mathcal{L}(g^{ab},R_{abcd})$ theory. Hence, these do not include the more general higher-derivative case. However, we are confident that Result \ref{ti1} extends as well to that case (especially in view of some available results for the Newtonian potential in some of these theories \cite{Modesto:2014eta,Giacchini:2016xns}).

%Let us estress again that the reason for considering a theory of the form $ \mathcal{L}(g^{ab},R_{abcd})$, instead of the more general \req{higherd}, is that the results used above for the Newtonian metric were obtained in \cite{PabloPablo,Aspects} only for the former subclass of theories.

%the spectrum and the Newtonian limit of these theories has been exhaustively studied \cite{PabloPablo,Aspects}, while a complete classification is still lacking for the general case $\mathcal{L}(g^{ab},R_{abcd},\nabla_{e}R_{abcd},\ldots)$. However, in view of the expression for the Newtonian potential in quadratic theories with an arbitrary number of derivatives \cite{Modesto:2014eta,Giacchini:2016xns}, we think it is very likely that the result applies also in the general case.

%Of course, our result can be rephrased as the fact that theories which include extra modes in their linearized spectrum do not admit static and spherically symmetric solutions of the form \req{Fmetric} representing the gravitational field of a spherical mass distribution.
% In the next subsection we illustrate this point in the simple case of $f(R)$ gravity.

 \subsection{``Unnaturality'' of Schwarzschild's black hole in $f(R)$ gravity}\label{unnatural}
 %{\bf Counter-example: $f( R )$}\\
Naively, the result found in the previous subsection seems to be incompatible with the fact that 
certain theories which propagate extra modes at the linearized level do also admit single-function black hole solutions of the form \req{Fmetric}. This apparent contradiction is not such. The reason is that, as we have stressed, our result holds only whenever \req{Fmetric} describes the gravitational field of a spherical mass distribution.

%, \ie when the corresponding solution is the \emph{natural} static and spherically symmetric black hole of the corresponding theory generalizing Einstein's gravity Schwarzschild-(A)dS solution in a non-trivial way, and reducing to it when the corresponding higher-order couplings are set to zero. 

In order to illustrate this point, let us consider the case of $f( R )$ gravity. It is well-known that this theory allows for the Schwarzschild-(A)dS solution in the absence of matter --- \eg \cite{delaCruzDombriz:2009et}. Hence, it possesses black hole solutions characterized by a single function. However, we also know that this theory propagates a scalar mode along with the massless graviton on the vacuum. Hence, our result implies that, even though the Schwarzschild-(A)dS metric is a vacuum solution of $f( R )$ gravity, it does not describe the external field of a generic spherically symmetric mass distribution for this theory. Let us verify this statement explicitly.  The $f( R )$ field equations coupled to matter read
 \begin{equation}\label{fr}
 	f'( R ) R_{ab}-\frac{1}{2}f( R )g_{ab}+\left(g_{ab}\Box-\nabla_{a}\nabla_{b}\right) f'( R )=\kappa T_{ab}\, ,
 \end{equation}
 where $\kappa$ is proportional to Newton's constant. Now, let us consider a static and spherically symmetric configuration with an energy-momentum tensor $T_{ab}$ such that $T_{ab}( r )=0$ if $r> r_0$, for certain $r_0$. %We allow the matter distribution to be discontinue at $r_0$. 
Further, let us assume this situation to be compatible with an exterior metric of constant scalar curvature, \ie satisfying  $R=\bar R$, where the constant $\bar R$ would be obtained from the algebraic equation $2\bar R f'( \bar R )-Df( \bar R )=0$.
If that was the case, \req{fr} would imply $R_{ab}\propto g_{ab}$, and we would obtain Schwarzschild-(A)dS in the exterior region. 

However, we will show that no constant-$R$ solution compatible with the above assumptions can exist in the outside region. 
The trace of the field equations reads
 \begin{equation}
 	(D-1)\Box f'( R )+R f'( R )-\frac{D}{2}f( R )=\kappa T\, .
 	\label{Requation}
 \end{equation}
 This can be thought of as an equation for $R$ (or $f'( R )$).
 %In this equation, $R$ (or $f'( R )$) can be thought as the variable. 
 Since we are considering a spherically symmetric situation, we can assume $R=R( r )$, which reduces \req{Requation} to an ordinary second-order differential equation for $R( r )$. A solution to this equation is then determined by specifying the values of $R$ and $dR/dr$ at some $r$.  
 %Let us impose the solution in the exterior region to be $R=\bar R=\rm{constant}$. This constant is determined by the algebraic equation $2\bar R f'( \bar R )-Df( \bar R )=0$.
 Now, in the transition point $r_0$ we must demand continuity and differentiability (otherwise there is no solution). Taking into account our assumptions for the exterior solution, this fixes the following boundary conditions for the internal one:
 \begin{equation}\label{bdc}
 R(r_0)=\bar R\, , \quad \frac{dR}{dr}(r_0)=0\, .
\end{equation} 
  The  solution for $r<r_0$ is then completely specified.  However, let us now consider a $(D-2)$-sphere of radius $r_s>r_0$ and unit normal $n^{a}$ at some time slice. Then, using the spherical symmetry of the problem and Stokes' theorem, it is straightforward to prove the following identities
 \begin{equation}
 	\Omega_{(D-2)} r_s^{D-2} \frac{d f'( R )}{dr}(r_s)=\oint_{S_{r_s}} d^{D-2}S\, n^{a}\nabla_{a}f'( R )=\int_{r<r_s} d^{D-1}x\sqrt{|g|}\Box f'( R )\, .
 \end{equation}
 Finally, taking into account that $f'( R )$ is constant for $r>r_0$ and using equation (\ref{Requation}), which holds in the $r<r_0$ region, we get
 \begin{equation}
 	\Omega_{(D-2)} r_s^{D-2} \frac{d f'( R )}{dr}(r_s)=\frac{1}{D-1}\int_{r<r_0} d^{D-1}x\sqrt{|g|}\left(\kappa T-R f'( R )+\frac{D}{2}f( R )\right).
	\label{contradiction}
 \end{equation}
 Now it is immediate to see that there is a problem here. Indeed, while the left-hand side (lhs) is zero, the rhs is non-vanishing in general. The vanishing of that integral could be imposed as a condition to the interior solution, and that would fix one integration constant. However, the interior solution is already completely specified by the boundary conditions \req{bdc}. We can see that those conditions do not depend on the particular form of the matter distribution, while the integral in the rhs does. Hence, that integral will be in general non-vanishing, which leads to a contradiction when compared to the lhs. This implies that no constant-$R$ solutions can describe the gravitational field in the outer region of a spherically symmetric matter distribution 
 for general $f(R)$ theories.\footnote{The fact that $f(R)$ theories do not allow for constant-$R$ solutions outside a source has been in fact known since long ago \cite{pechlaner1966,Michel:1973iu}. The problem is even worst, since it seems that the solution exterior to a source does not even exist in $f(R)$ gravity \cite{Mignemi:1991wa}. In this sense, $f(R)$ is a sick theory. Something similar could well happen for other higher-order gravities, so it is important to show that at least some of them do allow for solutions which can  describe the exterior field of a source.}
 
 Observe that in the Einstein gravity case, \ie when $f(R)=R-2\Lambda_0$, the contradiction in \req{contradiction} disappears, as the rhs vanishes in that case by virtue of Einstein's equation. This is naturally related to the absence of terms involving covariant derivatives of $f'(R)$ --- such as $\Box f'(R)$ --- in the equations of motion, which are in fact ultimately responsible for the appearance of the extra spin-0 mode in the linearized spectrum in the general $f(R)$ case.

 \section{Black holes from the on-shell action}\label{bho}
  In the previous section we showed that only theories sharing the linearized spectrum of Einstein gravity are susceptible of admitting single-function black holes of the form \req{Fmetric}
  describing the exterior gravitational field of spherical distributions of matter. In this section we prove Result \ref{theo}, which provides in turn a sufficient condition for identifying such theories, as well as a simple method for determining the equation satisfied by $f(r)$ in each case. The proof of this result gives rise to a simple method for constructing higher-derivative gravities satisfying the hypothesis of Result \ref{theo}, as well as for obtaining the equation that determines $f(r)$ in each case. We detail this procedure in section \ref{recipe}. In sections \ref{quadratic} and \ref{cubic}, we apply it to the cases of $D$-dimensional quadratic and cubic gravities respectively, which serves as an illustration of the method, and provides further evidence for Conjectures \ref{conj} and \ref{conji2} .

 \subsection{Sufficient condition for single-function solutions}\label{bho1}

%\commentt{bla bla}
% Note that the existence of solutions of the form \req{Fmetric} does not guarantee that any of them must necessarily correspond to a black hole. Whether or not this is the case for a given theory requires further analysis, but we will... \ref{conj} \commentt{finish this}
   
 % Later, we will argue that, indeed, all the theories which allow for physically reasonable black holes with a single function are of this form. \commentt{rewrite a bit this paragraph}
 
%Let us start by mentioning that while the results in the previous section are valid for general $\mathcal{L}($Riemann$)$ theories, here we consider an even more general case, namely, 
%we also allow for an arbitrary dependence on terms involving covariant derivatives of the Riemann tensor, \ie\

In this section we consider the general action \req{higherd}, \ie with respect to the previous section, we also allow for an arbitrary dependence on terms involving covariant derivatives of the Riemann tensor.
%as opposed to the $\mathcal{L}($Riemann$)$ subclass,  which we sometimes call \emph{higher-order} gravities.}
% Let us consider the most general higher-derivative gravity, with Lagrangian density $\mathcal{L}(g^{\mu\nu},R_{\mu\nu\rho\sigma},\nabla_{\alpha}R_{\mu\nu\rho\sigma},\ldots)$.
 Then, as explained before, if the Lagrangian contains up to $n$ derivatives of the Riemann tensor, the field equations generally involve $2n+4$ derivatives of the metric.  
Such equations can be equivalently studied, for a general static and spherically symmetric ansatz of the form \req{Nf},  by considering the action functional $S[N,f]$, where all the metrics in the gravitational Lagrangian are evaluated on that ansatz, \ie
\begin{equation}
	S[N,f]=\Omega_{(D-2)}\int dt \int dr L_{N,f}\, ,
\end{equation}
where $L_{N,f}$ was defined in \req{LNf}.
%\begin{equation}
%L_{N,f}(r, N( r ), N'( r ), N''( r ),\ldots, f( r ), f'( r ), f''( r ),\ldots)\equiv N( r )r^{D-2}\mathcal{L}\big|_{g^{ab}=g^{ab}_{N,f}}\, ,
%\end{equation}
 %is an effective Lagrangian for $N$ and $f$, and $\Omega_{(D-2)}$ is the area of $S^{(D-2)}$. 
 %Now, observe that the action must be invariant under rescalings $N\rightarrow N \alpha(t)$ for an arbitrary function $\alpha(t)$, since they are equivalent to a time reparametrization. In particular,
More explicitly, the variations of $S[N,f]$ with respect to $N$ and $f$ are related to the $tt$ and $rr$ components of the corresponding field equations $\mathcal{E}_{ab}\equiv \frac{1}{\sqrt{ |g|}}\frac{\delta S}{\delta g^{ab}}$ through \cite{PabloPablo2}
 \begin{equation}
 \frac{1}{\Omega_{(D-2)}r^{D-2}}\frac{\delta S[N,f]}{\delta N}=\frac{2\mathcal{E}_{tt}}{ f N^2}\, , \quad \frac{1}{\Omega_{(D-2)}r^{D-2}}\frac{\delta S[N,f]}{\delta f}=\frac{\mathcal{E}_{tt}}{N f^2}+N \mathcal{E}_{rr}\, .
 \end{equation}
 Hence, imposing the Euler-Lagrange equations of $N$ and $f$ to hold is equivalent to imposing $\mathcal{E}_{tt}=\mathcal{E}_{rr}=0$. Finally, the Bianchi identity $\nabla^a \mathcal{E}_{ab}=0$ automatically makes the angular components vanish whenever  $\mathcal{E}_{tt}=\mathcal{E}_{rr}=0$.

 Observe now that the constant rescaling $N\rightarrow N\alpha$, for an arbitrary $\alpha$, is equivalent to the time rescaling $t\rightarrow t\, \alpha$, which leads to the following identities
 %This way, we are able to deduce the following equality
\begin{equation}
	S[\alpha N, f]=\Omega_{(D-2)}\int dt \int dr L_{\alpha N,f}=\Omega_{(D-2)}\int d(\alpha t) \int dr L_{N,f}
	%=\alpha\Omega_{(D-2)}\int dt \int dr L_{N,f}
	=\alpha S[N,f]\, .
\end{equation}
This implies that both $S[N,f]$ and $L_{N,f}$ are homogeneous of degree 1 in $N$. Moreover, the Lagrangian is formed by products, quotients and derivatives of the metric components, and the only homogeneous monomials of degree 1 which can be formed in this way with $N$ and its derivatives are of the form $N^{i_1}N'^{i_2}N''^{i_3}\cdots(N^{(n+2)})^{i_{n+3}}$, with $i_k$ integers such that $i_1+\ldots +i_{n+3}=1$. Also, we must have $i_k\ge0$ for $k>1$ because the derivatives cannot appear in the denominator. Taking this into account, we observe that the Lagrangian can always be expanded in the following way,\footnote{$N^{(i)}$ stands for the $i$-th derivative of $N$ with respect to $r$, and so on.}
\begin{equation}
	L_{N,f}=N L_{f}+\sum_{i=1}^{n+2}N^{(i)}F_i+\mathcal{O}(N'^2/N)\, ,
	\label{homogeneus}
\end{equation}
where 	$L_f(r, f, f', f'',\ldots)\equiv L_{N=1,f}$ is the effective Lagrangian resulting from the evaluation of the gravitational Lagrangian in the single-function ansatz (\ref{Fmetric}), and the $F_i=F_i(r,f, f', f'',\ldots, f^{(n+2)})$ are functions of $f$ and its derivatives. Finally, $\mathcal{O}(N'^2/N)$ denotes all the terms which are at least quadratic  in derivatives of $N$, 
\begin{equation}\label{ooo}
	\mathcal{O}(N'^2/N)\equiv \sum_{i,j=1}^{n+2}\frac{N^{(i)}N^{(j)}}{N}F_{ij}+\sum_{i,j,k=1}^{n+2}\frac{N^{(i)}N^{(j)}N^{(k)}}{N^2}F_{ijk}+\ldots\, ,
\end{equation}
where, again, $F_{ij}$, $F_{ijk}$, etc., only depend on $f$ and its derivatives.
%Also, we have taken into account that $L_{N=1,f}=L_f$, so that is why the term proportional to $N$ is $L_f$ in (\ref{homogeneus}). 

The analysis so far is completely general. Let us now make use of the hypothesis of Result \ref{theo}: we assume that \req{EELL} holds, \ie that the Euler-Lagrange equation of $f(r)$ for the Lagrangian $L_f$ vanishes identically. Of course, this is equivalent to the assumption that $L_f$ is a total derivative, this is, that there exists a function $F_0(r, f, f',\ldots, f^{(n+1)})$ such that
\begin{equation}
	L_f=F_0'\,,
\end{equation}
where again the prime denotes a total derivative with respect to $r$. Using this result and the expansion (\ref{homogeneus}), we can express the action functional as
\begin{equation}\label{acc}
	S[N,f]=\Omega_{(D-2)}\int dt \int dr \left[ N \left(F_0+\sum_{i=1}^{n+1}(-1)^iF_ i^{(i-1)}\right)'+\mathcal{O}(N'^2/N)\right]\, ,
\end{equation}
where we have integrated by parts several times. Now we are ready to compute the Euler-Lagrange equations for $N$ and $f$. First, it is immediate to see that the equation $\delta_f S=0$ is trivially satisfied whenever $N'=0$. Hence, we can just set $N$ to a constant value, which we choose to be one. On the other hand, variation with respect to $N$ and evaluation at $N=1$ yields
\begin{equation}
	\delta_N S=\Omega_{(D-2)}\int dt \int dr  \delta N \left(F_0+\sum_{i=1}^{n+1}(-1)^iF_ i^{(i-1)}\right)'=0\, .
\end{equation}
Hence, integrating once the equation, we get
\begin{equation}
	F_0+\sum_{i=1}^{n+1}(-1)^iF_ i^{(i-1)}=C\, ,
	\label{fequation}
\end{equation}
for some integration constant $C$. This is the differential equation which determines $f(r)$ in each case. In order to determine its order, let us consider the Bianchi identity, $\nabla_{a}\mathcal{E}^{ab}=0$. The $\nu=r$  component reads
\begin{equation}
	\frac{ d \mathcal{E}^{rr}}{dr}+\left(\frac{2}{r}-\frac{1}{2}f^{-1}f'\right)\mathcal{E}^{rr}+\frac{1}{2}f f'\mathcal{E}^{tt}-r f\mathcal{E}^{\theta\theta}-\sin^2(\theta) r f\mathcal{E}^{\varphi\varphi}=0\, .
\end{equation}
Since all the components of $\mathcal{E}^{ab}$ contain derivatives up to order $2n+4$ and this identity relates the derivative of $\mathcal{E}^{rr}$ to the rest of components (without derivatives), we must conclude that in fact $\mathcal{E}^{rr}$ contains derivatives up to order $2n+3$.
Now, in \req{fequation} we have integrated the equation once, so the order of the equation is reduced yet another order. Therefore, \req{fequation} is in general of order $2n+2$, which means two orders less than the equations determining $N(r)$ and $f(r)$ in the general case. Naturally, in $ \mathcal{L}(g^{ab},R_{abcd})$ theories, for which $n=0$,  (\ref{fequation}) becomes a differential equation of order 2 or less --- see \req{fff} below. This completes the proof of the first part of Result \ref{theo}.

Let us now proceed with the second.
%{\bf (2)} In order to continue, let us introduce the functional derivative $\frac{\delta S}{\delta N}$ as
%\begin{equation}
%	\delta_N S=\Omega_{(D-2)}\int dt \int dr  \delta N\frac{\delta S}{\delta N},
%\end{equation}
%and the same for the variation with respect to $f$.
In order to do so, let us start by adding some minimally coupled matter to the gravity action \req{higherd}, $S\rightarrow S + S_{\ssc \rm matter}$, where $S_{\ssc \rm matter}= \int d^Dx \sqrt{|g|}\, L_{\ssc \rm matter} $. The field equations would read now $\mathcal{E}_{ab}=\frac{1}{2}T_{ab}$, where the matter stress-energy tensor is defined as usual,
$
	T_{ab}=-\frac{2}{\sqrt{|g|}}\frac{\delta S_{\ssc \rm matter}}{\delta g^{ab}}.
$
Then, it can be shown that the equations for $f$ and $N$ corresponding to the general ansatz \req{Nf} read
\begin{equation}
\left(F_0+\sum_{i=1}^{n+1}(-1)^iF_ i^{(i-1)}\right)'+\mathcal{O}(N')=r^{D-2} f N^2T^{tt}, \quad \frac{\delta S[N,f]}{\delta f}=\Omega_{(D-2)}\frac{r^{D-2}}{2}\left(N^3T^{tt}+\frac{N}{f^2}T^{rr}\right)\, .
	\label{matterequations}
\end{equation}
%where we have introduced the notation $\mathcal{F}=F_0+\sum_{i=1}^{n+1}(-1)^iF_ i^{(i-1)}$.
We are interested in a compact, spherically symmetric and static source of radius $r_0$, so that $T_{ab}( r )=0$ if $r>r_0$. 
%Now, using the hypothesis of Result \ref{theo}, we assume that for $r>r_0$, the solution is described by a single function \req{Fmetric}, so that \req{fequation} holds in that region for some arbitrary constant $C$.
%Outside the source we have a solution with $N=1$ and $\mathcal{F}=C$, for an arbitrary constant $C$, which corresponds to a vacuum solution of the field equations.
Let us begin by constructing the interior solution, $r<r_0$.
In order to find it, we need to impose some boundary conditions at $r=r_0$. In particular, since we want an exterior solution with $N=1$, we set $N(r_0)=1$, $N'( r_0)=0, \ldots, N^{(n_{\rm{max}}-1)}( r_0 )=0$, where $n_{\rm{max}}$ is the highest derivative of $N$ in the equations. On the other hand, we allow the values of $f$ and its derivatives at $r=r_0$ to be arbitrary, but observe that once they are specified, the interior solution is fully determined. The idea is now to find the exterior solution $r>r_0$ which corresponds to this internal configuration. In order to do so, note that the second equation in \req{matterequations} is solved outside the source for $N=1$, and this choice guarantees the continuity of $N$ and its $n_{\rm{max}}-1$ first derivatives at $r=r_0$.
On the other hand, applying Stokes' theorem on the first equation in \req{matterequations} yields
\begin{equation}
F_0(r)+\sum_{i=1}^{n+1}(-1)^iF_ i^{(i-1)}(r)=\frac{1}{\Omega_{(D-2)}}\int_{r<r_0}d^{D-1}x\sqrt{|g|}\left[f N T^{tt}-\frac{\mathcal{O}(N')}{N r^{D-2}}\right]\, ,
\label{FequationT}
\end{equation}
where the lhs is evaluated on arbitrary $r>r_0$ while the rhs is an integral over the matter distribution, and thus, it is independent of $r$. Therefore, \req{fequation} holds outside the source but now the constant $C$ is not arbitrary, but determined by the mass distribution.\footnote{Note that \req{FequationT} and (\ref{contradiction}) are similar, but there is a conceptual difference. In (\ref{contradiction}) we were forcing the rhs to be zero, and this was the origin of the contradiction found there, while \req{FequationT} is simply determining the value of the constant $C$, so no contradiction is found.} Even though it seems that $C$ could depend on the density profile or on the radius of the source, as we show next, studying the asymptotic behavior of the solution one finds that it is always proportional to the total mass.  Indeed, assuming that our higher-derivative theory has a well-defined Einstein gravity limit as in \req{higherd},  then the asymptotic expansion of \req{fequation} becomes
\begin{equation}
\frac{1}{16\pi G}\left[-\frac{2}{(D-1)}\Lambda_0 r^{D-1}-(D-2)(f-1)r^{D-3}+\ldots\right]=C\,.
\end{equation}
In the asymptotically flat case, $\Lambda_0=0$, this equation implies
\begin{equation}
f( r )=1-\frac{16 \pi G C}{(D-2) r^{D-3}}+\mathcal{O}(r^{2-D})\, .
\end{equation}
Now, for an asymptotically flat spacetime, no matter the higher-derivative corrections, the total mass can be computed according to ADM formula \cite{Arnowitt:1960es,Arnowitt:1960zzc,Arnowitt:1961zz,Deser:2002jk}, which yields 
\begin{equation}
M=\frac{(D-2)\Omega_{(D-2)}}{16\pi G}\lim_{r\rightarrow\infty} r^{D-3}\left(\frac{1}{f(r)}-1\right)\, .
\label{massC}
\end{equation}
Then, we obtain
\begin{equation}
C=\frac{M}{\Omega_{(D-2)}}\, .
\label{Cmass}
\end{equation}
In the case of a non-vanishing cosmological constant, this relation still holds. Indeed, in the asymptotically (A)dS case, the mass is given by a generalized version of \req{massC} \cite{Abbott:1981ff,Deser:2002jk}, but the identity above remains unchanged.
%Intuitively, the constant $C$ can be related to a local source which is independent of the asymptotic behavior of the solution, so one expects that it is the same in all cases. More rigorously, one can compute explicitly the mass of the solution in the asymptotically (A)dS case in several examples and check that indeed \req{Cmass} is correct. One has to take into account that in the asymptotically (A)dS case, the mass is given by a generalized version of \req{massC} \cite{Abbott:1981ff,Deser:2002jk}. 
In sum, the equation for $f$ can always be written as
\begin{equation}
F_0-F_1+F_2'-\ldots=\frac{M }{\Omega_{(D-2)}}\, .
\end{equation}
Returning to the interior solution, remember that we intentionally left undetermined the values of $f$ and their first derivatives at $r=r_0$. Then, these values are chosen so that the interior solution is well glued with the exterior one (\ie there is continuity and differentiability). 

This completes the proof that the vacuum solutions of the form \req{Fmetric} do describe the exterior gravitational field of spherically symmetric matter distributions and that they are characterized by the total mass.
There is no unicity though, since the vacuum equation \req{fequation} could have more than one solution even for a fixed $M$. In other words, the solutions which represent the exterior field of a source could have ``hair''. When we impose the solution to be a black hole, this is no longer the case, as the existence of a horizon imposes a natural boundary condition which kills the possible hair, as we explicitly showed for $D=4$ Einsteinian cubic gravity in \cite{PabloPablo2} --- see also Section \ref{planar} for a new five-dimensional example.\footnote{Further evidence was recently provided in \cite{Hennigar:2017ego} for the asymptotically flat black holes of $D$-dimensional Generalized quasitopological gravity --- see also Section \ref{cubic}.}
This result is captured in Conjeture \ref{conj}.

The third part of Result \ref{theo} follows from the first two along with Result \ref{ti1}, which we proved in the previous section.

Let us conclude this subsection by stressing that Lovelock theories, Quasitopological gravity (and its higher-order generalizations), four-dimensional Einsteinian cubic gravity and Generalized quasitopological gravity satisfy the hypothesis of Result \ref{theo}. On the one hand, our results explain why they all have the same linearized equations as Einstein gravity (up to a redefinition of Newton's constant \cite{Aspects}). On the other, they imply that their static and spherically symmetric black hole solutions of the form \req{Fmetric} are the natural generalizations of Schwarzschild-(A)dS, as they represent the exterior field of spherically symmetric mass distributions.

 % in view of the results the results of the following sections suggest a remarkable conjecture:\\
%  \begin{conjecture}\label{conj} For all the theories fulfilling the hypothesis of Result \ref{theo}: 
%
 %\begin{enumerate}
 %	\item{(No-hair) Given a mass $M$, there is a unique solution of the form (\ref{Fmetric}) which represents a black hole of that mass}
 %	\item{The thermodynamics of this black hole can be determined by solving a system of algebraic equations with no free parameters}
 %\end{enumerate}
%\end{conjecture}

\subsection{A recipe}\label{recipe}
The results obtained in the previous subsection provide a very simple and efficient method for identifying higher-derivative gravities with simple black hole solutions of the form \req{Fmetric}, and for characterizing those solutions. For the same price, the solutions constructed in this way correspond to the exterior field of a spherically symmetric mass distribution, and the corresponding theories are automatically equivalent to Einstein gravity at the linearized level. Our method is a refinement of an often utilized procedure, \eg in \cite{Palais:1979rca,Deser:2003up,Quasi,PabloPablo2,Dehghani:2011vu,Cisterna:2017umf,Hennigar:2017ego}, consisting in evaluating $L_{N,f}$ and performing a repeated integration by parts in the aim of bringing it to the form \req{acc} for a particular combination of couplings. 

Let us now present our method in the form of a recipe ready to be applied to any higher-derivative gravity. From a computational perspective, our procedure is considerably faster than the one just described. It involves two trivial on-shell evaluations of the higher-derivative action, computing the Euler-Lagrange equation of a one-dimensional Lagrangian, writing an expression as a total derivative of another function (which is guaranteed to be possible), and computing some trivial derivatives. Here is the recipe:

\begin{enumerate}
	\item Evaluate the gravity Lagrangian on the single-function ansatz \req{Fmetric}, namely, $L_f(r,f',f'',\ldots)\equiv r^{D-2}\mathcal{L}|_{g^{ab}=g^{ab}_f}$.
	\item Compute the Euler-Lagrange equation of $f(r)$ for the effective Lagrangian $L_f(r,f',f'',\ldots)$.
	\item Fix the higher-derivative couplings in a way such that this equation is identically satisfied, \ie impose $\delta L_f/\delta f=0$ for all $f(r)$.
	\item Find $F_0$, namely, the function of $f(r)$ satisfying $L_f=F_0'$. 
	\item Substitute the general ansatz \req{Nf} in the corresponding gravity Lagrangian: $L_{N,f}(r,f, N,f', N', \ldots)$ $\equiv N( r )r^{D-2}\mathcal{L}\big|_{g^{ab}=g^{ab}_{N,f}}$. The result should take the form \req{homogeneus}, where now $L_f=F_0'$. 
	\item Identify the functions $F_i$ by inspection. 
	\item Plug $F_0$ and the (corresponding derivatives of the) $F_i$ in \req{fequation}. This is the equation that determines $f(r)$.
\end{enumerate}
The first three steps select, from all the possible theories considered originally, the ones which allow for single-function solutions of the form \req{Fmetric}. The last four allow one to determine the differential (or algebraic) equation which needs to be solved in order to determine $f(r)$ for the corresponding theory.

%The procedure outlined above turns out to be considerably faster
%Computationally, our method is considerably faster, as it only involves obtaining the Euler-Lagrange equations for $L_f$ 

\subsubsection{Quadratic gravities}\label{quadratic}
In order to illustrate this method, let us apply it to the $D$-dimensional quadratic theory
%\subsection{Quadratic theories in general dimensions}
%\begin{equation}
%\mathcal{L}_{\text{quadratic}}=\frac{1}{16\pi G_{\rm \ssc N}} \left[-2\Lambda_0+R+G_{\rm \ssc N}^{\frac{2}{D-2}}\sum_{i=1}^3 \alpha_i \mathcal{L}_{i}^{(2)} \right]
%\end{equation}
\begin{equation}
\mathcal{L}_{\rm \ssc quadratic}=\frac{1}{16\pi G}\left[-2\Lambda_0+R+ \alpha_1 R^2+\alpha_2 R_{ab}R^{ab}+\alpha_3 R_{abcd}R^{abcd} \right]\, .
\end{equation}
%\begin{table}[hpt] 
%	\begin{center}
%		\hspace*{-1cm}
%		\begin{tabular}{c| c| c}
%			%\hline
%			$\mathcal{L}^{(2)}_1=R^2$ &  $\mathcal{L}^{(2)}_2=R_{ab}R^{ab}$& $\mathcal{L}^{(2)}_3=R_{abcd}R^{abcd}$ \\
%			%\hline
%			%$m_s^2=+\infty$& Critical  & No dynamical scalar & Einstein-like  \\
			%\hline
%		\end{tabular}
%		\hspace*{-1cm}
		%\caption{}
%		\label{tablaa}
%	\end{center}
%\end{table}
%\begin{align}
%R_f&=(D-2)\frac{(D-3)(1-f)-2r f'}{r^2}-f''\, , \\
%\left(R_{ab}R^{ab}\right)_f&=(D-2)\frac{2(D-3)^2(-1+f)^2+4(D-3)r(f-1)f'+D r^2 f'^2+2r^3 f' f''}{2r^4}+\frac{f''^2}{2}\, , \\
%\left(R_{abcd}R^{abcd}\right)_f&=2(D-2)\frac{(D-3)(-1+f)^2+ r^2 f'^2}{r^4}+f''^2\, ,
%\end{align}
Evaluating the Lagrangian on the single-function metric ansatz, we obtain the effective Lagrangian
\begin{equation}
\begin{aligned}
L_f=&\frac{1}{16\pi G}\Big[-2\Lambda_0 r^{D-2}+(D-2)(D-3) r^{D-4} (f-1)+2(D-2)r^{D-3}f'+r^{D-2}f''\\
&+\alpha_1 r^{D-6}\left((D-2)(D-3)(f-1)+2(D-2)rf'+r^2f''\right)^2\\
&+\alpha_2r^{D-6}\left((D-2)((D-3)(f-1)+rf')+((D-2)rf'+r^2f'')^2/2\right)\\
&+\alpha_3r^{D-6}\left(2(D-2)(D-3)(f-1)^2+2(D-2)r^2f'^2+r^4 f''^4\right)\Big].
\end{aligned}
\end{equation}
From this, it is straightforward to compute the Euler-Lagrange derivative, which yields
\begin{equation}
\begin{aligned}
\frac{\delta L_f}{\delta f}=&\frac{(D-2)}{16\pi G}\left[ (3\alpha_1+\alpha_2+\alpha_3)\left(4(D-3)(f-1)-2r^2f''\right)\right.\\
&\left.+(2\alpha_1+\alpha_2+2\alpha_3)\left((D-4)r^2 f''+2r^2f^{(3)}+r^4 f^{(4)}\right)\right]\,.
\end{aligned}
\end{equation}
 Then, applying the third step of the recipe, we find that
imposing $\delta L_f/ \delta f= 0$ $ \forall f(r)$ fixes $\alpha_1=\alpha_3=-\alpha_2/4=\alpha$ which, unsurprisingly, leads to the usual Gauss-Bonnet combination. Having fixed these couplings, we can compute $F_0$ for this theory, which turns out to read
\begin{align}\notag
16\pi GF_0&=(D-2)r^{D-3}(1-2r^2\Lambda_0/((D-2)(D-1))-f)+f'(2(D-3)(D-2) r^{D-4}(f-1)\alpha-r^{D-2}) \\ 
&+(D-4)(D-3)(D-2)r^{D-5} (f-1)^2\alpha\, .
\end{align}
The next step is to evaluate the Lagrangian in the general metric ansatz with two functions. Amusingly, the effective Lagrangian $L_{N,f}$, which in general takes the form \req{homogeneus}, does not contain any $\mathcal{O}(N'^2/N)$ term in this case, and is simply given by \req{siss},
where
\begin{align}\notag
16\pi GF_1&=r^{D-5}(-3f' r^3+2(D-3)(D-2)(5f-3)f' r \alpha-2(D-2)f (r^2-2(D-4)(D-3)(f-1)\alpha))\, ,\\
16\pi GF_2&=-2r^{D-4}f(r^2+2(D-3)(D-2)\alpha(1-f))\, .
\end{align}
This is all we need to determine the equation of $f(r)$, \req{Fmetric}, which in this case (and for any $\mathcal{L}(g^{ab},R_{abcd})$ theory satisfying the hypothesis of Result \ref{theo}) reads: 
\begin{equation}\label{fff}
 F_0-F_1+F_2'=C\,.
\end{equation}
Explicitly, one finds 
\begin{equation}\label{GB}
r^{D-3}\left((D-2)(1-f)-\frac{2r^2\Lambda_0}{(D-1)}\right)+\alpha (D-4)(D-3)(D-2) r^{D-5}(f-1)^2=\frac{16\pi G M}{\Omega_{(D-2)}}\,,
\end{equation}
where we have taken into account \req{Cmass}. This equation can be solved for $f(r)$ to yield
\begin{equation}
f(r)=1+\frac{r^2}{2(D-3)(D-4)\alpha}\left[1\mp \sqrt{1+\frac{8\alpha \Lambda_0 (D-4)(D-3)}{(D-2)(D-1)}+\frac{64 \alpha\pi G M(D-3)(D-4)}{(D-2)\Omega_{(D-2)}r^{D-1}}} \right]\,.
\end{equation}
This is naturally the blackening factor of the usual $D$-dimensional static and spherically-symmetric Gauss-Bonnet black hole.\footnote{Observe that for $D=4$ and $D=3$, it reduces to the usual Schwarzschild-(A)dS solution.} 

Notice that the fact that \req{GB} is algebraic instead of a second-order differential equation is a non-generic feature which also occurs for general Lovelock \cite{Wheeler:1985nh,Wheeler:1985qd,Boulware:1985wk,Cai:2001dz,Dehghani:2009zzb,deBoer:2009gx} and Quasitopological theories \cite{Quasi,Quasi2,Dehghani:2011vu,Cisterna:2017umf}, but not in other cases like Einsteinian cubic gravity \cite{PabloPablo2} or the recently constructed Generalized quasitopological gravity \cite{Hennigar:2017ego} --- see below. Interestingly, the algebraicity of \req{fff} (which is a result of the non-trivial cancellation of the different terms in $F_0$, $F_1$ and $F_2'$ involving derivatives of $f(r)$)  appears to be related to the absence of $\mathcal{O}(N'^2/N)$ terms in $L_{N,f}$. This is, very likely, a general feature which we encapsule in Conjecture \ref{conji2}. 

In addition, let us comment that by using the solution above and the results in \cite{Deser:2002jk} it is easy to check explicitly that $M$ is in fact the total mass in any dimension and for any asymptotic behavior.

\subsubsection{Cubic gravities}\label{cubic}
Let us be less detailed with the cubic case. If the analysis is repeated at this order in curvature, one is led to the recently constructed Generalized quasitopological gravity \cite{Hennigar:2017ego}, whose action is given by
\begin{equation}\label{gQuasi2}
\mathcal{L}_{\rm \ssc cubic}=\frac{1}{16\pi G} \left[-2\Lambda_0+R+\alpha \mathcal{X}_4+ \gamma \mathcal{X}_6+\mu \mathcal{Z}_D+\xi \mathcal{S}_D\right]\, .
\end{equation}
In this expression, $\mathcal{X}_6$ is the cubic Lovelock density \cite{Lovelock1,Lovelock2}, $\mathcal{Z}_D$ is the $D$-dimensional cubic quasitopological term \cite{Quasi,Quasi2}, and $\mathcal{S}_D$ is the new term \cite{Hennigar:2017ego} which generalizes Einsteinian cubic gravity \cite{PabloPablo} to higher-dimensions.\footnote{\label{ECGf}Let us stress that Einsteinian cubic gravity (with its characteristic cubic term \req{ecg}) was in fact defined for general dimensions in \cite{PabloPablo}, where it was shown that it is the most general dimension-independent theory which, up to cubic order, shares the linearized spectrum of Einstein gravity. Hence, $\mathcal{S}_D$ is a generalization of the $D=4$ version of the Einsteinian cubic gravity term \req{ecg}  which, as opposed to its $D\geq 5$ versions, satisfies the hypothesis of Result \ref{theo}.} This term is explicitly given by
\begin{equation}
\begin{aligned}
\mathcal{S}_D=&14\tensor{R}{_{a}^{c}_{b}^{d}}\tensor{R}{_{c}^{e}_{d}^{f}}\tensor{R}{_{e}^{a}_{f}^{b}} +2\tensor{R}{_{abcd}}\tensor{R}{^{abc}_{e}}R^{de}-\frac{(38-29 D+4 D^2)}{4(D-2)(2D-1)}\tensor{R}{_{abcd}}\tensor{R}{^{abcd}}R\\
&-\frac{2(-30+9D+4D^2)}{(D-2)(2D-1)}\tensor{R}{_{abcd}}\tensor{R}{^{ac}}\tensor{R}{^{bd}}
-\frac{4(66-35D+2D^2))}{3(D-2)(2D-1)} R^{b}_{a} R_{b}^{c} R_{c}^{a}\\
&+\frac{(34-21D+4D^2)}{(D-2)(2D-1)}R_{ab}R^{ab} R-\frac{(30-13D+4D^2)}{12(D-2)(2D-1)} R^3\, .
\end{aligned}
\label{SD}
\end{equation}
When restricted to four dimensions, one finds that \cite{Hennigar:2017ego} 
\begin{equation}
\mathcal{S}_4=\mathcal{P}-\frac{1}{4}\mathcal{X}_6+4\mathcal{C}\, ,
\end{equation}
where
\begin{equation}\label{ecg}
\mathcal{P}=12 \tensor{R}{_{a}^{c}_{b}^{d}}\tensor{R}{_{c}^{e}_{d}^{f}}\tensor{R}{_{e}^{a}_{f}^{b}}+\tensor{R}{_{ab}^{cd}}
\tensor{R}{_{cd}^{ef}}\tensor{R}{_{ef}^{ab}}-12\tensor{R}{_{abcd}}\tensor{R}{^{ac}}\tensor{R}{^{bd}}+8R^{b}_{a} R_{b}^{c} R_{c}^{a}\, ,
\end{equation}
is the usual Einsteinian cubic gravity term \cite{PabloPablo}, $\mathcal{X}_6$ identically vanishes (for any metric in $D\leq 5$), and $\mathcal{C}$ is a cubic term which does not contribute to the equations of motion for a static and spherically symmetric ansatz \req{Nf}. It is in this sense that $\mathcal{S}_D$ provides a $D$-dimensional generalization of four-dimensional Einsteinian cubic gravity, as explained in \cite{Hennigar:2017ego}. As emphasized in that paper, the linearized spectrum of \req{gQuasi2} coincides with Einstein's, which of course can be now understood as a consequence of this theory satisfying the hypothesis of Result \ref{theo}. This also implies that the black hole solutions of \req{gQuasi2} constructed solving the corresponding equation for $f(r)$ \cite{Hennigar:2017ego} describe the exterior gravitational field of generic spherically symmetric distributions of mass for the theory \req{gQuasi2} minimally coupled to matter.

We also point out that in the absence of $\mathcal{S}_D$ (or, equivalently, $\mathcal{P}$ in $D=4$) the equation that determines $f(r)$ is algebraic, and no $\mathcal{O}(N'^2/N)$ terms of the form \req{ooo} appear in $L_{N,f}$. This is no longer the case when $\mathcal{S}_D$ is included. These observations strongly support Conjecture \ref{conji2}.
Finally, let us also mention that for all the different theories contained in \req{gQuasi2}, Conjecture \ref{conj} holds.

\section{Planar black holes in $D=5$ Generalized quasitopological gravity}\label{planar}
In this section we study the asymptotically AdS$_5$ planar black holes of Generalized quasitopological gravity \cite{Hennigar:2017ego}. As explained in the introduction, this will serve three main purposes: i) showing that many of the results developed in this paper can be equally applied to obtain flat and hyperbolic  black hole solutions, ii) providing further support for Conjecture \ref{conj}, iii) paving the way for future holographic studies of $D\geq 5$ Generalized quasitopological gravity \cite{Hennigar:2017ego} and $D=4$ Einsteinian cubic gravity \cite{PabloPablo}, \eg along the lines of \cite{Brigante:2007nu,deBoer:2009gx,Buchel:2009sk,Myers:2010jv,Myers:2010tj}. Let us slightly adapt our conventions \cite{Quasi} and consider the following cubic theory,
%\begin{equation}
%S=\frac{1}{16\pi G_5}\int d^5x\sqrt{|g|}\left[\frac{12}{L^2}+R+\frac{\alpha L^2}{2} \mathcal{X}_4+\frac{7\lambda L^4}{4} \mathcal{Z}_5-\frac{\zeta L^4}{560} \left(81\mathcal{S}_5+1106 \mathcal{Z}_5\right)\right],
%\end{equation}
\begin{equation}\label{5dsg}
S=\frac{1}{16\pi G_5}\int d^5x\sqrt{|g|}\left[\frac{12}{L^2}+R+\frac{\alpha L^2}{2} \mathcal{X}_4+\frac{7\mu L^4}{4}  \mathcal{Z}_5-\frac{81\xi L^4}{632} \mathcal{S}_5\right]\, ,
\end{equation}
where 
we chose the cosmological constant to be negative and given by $\Lambda_0=-6/L^2$, where the scale $L$ coincides with the AdS$_5$ radius when $\mu=\xi=0$. Here,
%$\mathcal{X}_4=R^2-4R_{ab}R^{ab}+R_{abcd}R^{abcd}$ is the Gauss-Bonnet density,
\begin{equation}
\begin{aligned}
\mathcal{Z}_5&=\tensor{R}{_{a}^{c}_{b}^{d}}\tensor{R}{_{c}^{e}_{d}^{f}}\tensor{R}{_{e}^{a}_{f}^{b}}+\frac{1}{56}\Big(-72\tensor{R}{_{abcd}}\tensor{R}{^{abc}_{e}}R^{de}+21\tensor{R}{_{abcd}}\tensor{R}{^{abcd}}R+120\tensor{R}{_{abcd}}\tensor{R}{^{ac}}\tensor{R}{^{bd}}\\
&+144 R^{b}_{a} R_{b}^{c} R_{c}^{a}- 132R_{ab}R^{ab}R+15 R^3\Big)\, ,
\end{aligned}
\end{equation}
is the quasitopological term \cite{Quasi2,Quasi} and $\mathcal{S}_5$ is the new curvature-cubed interaction \req{SD} in five dimensions \cite{Hennigar:2017ego}.
%We have introduced the combination $81\mathcal{S}_5+1106 \mathcal{Z}_5$ for convenience, as it simplifies the equations.
%In the preceding sections we have considered the case of black holes with spherical topology but now we will study the planar case. The formulas for general horizon topology and cosmological constant can be found in the appendix. 
The metric ansatz can now be written as
\begin{equation}
ds^2=\frac{r^2}{L^2}\left(-N( r )^2 f( r )dt^2+dx^2+dy^2+dz^2\right)+\frac{L^2}{r^2f(r)} dr^2\, .
\end{equation}
As expected, \req{5dsg} allows for solutions with constant $N$. The second-order differential equation which determines $f( r )$ reads
\begin{equation}
1-f+\alpha f^2+(\mu+\xi) f^3+\tilde{\xi} \left[-\frac{ r^3}{2}f'^3+\frac{15r^2}{4}ff'^2+\frac{3r^3}{2}ff'f''\right]=\frac{\omega^4}{r^4}\, ,
\label{5Dequation}
\end{equation}
where $\omega$ is an integration constant and where we introduced the constant $\tilde{\xi}\equiv 70\xi/79$ for clarity reasons.
Unlike the cases of Lovelock and Quasitopological gravities, this equation is differential instead of algebraic (and $\mathcal{O}(N'^2/N)$ terms of the form \req{ooo} appear in $L_{N,f}$, in agreement with Conjecture \ref{conji2}), which means that there could exist several families of solutions. However, let us show that there is in fact a unique solution representing an asymptotically AdS black hole. First, we demand that  $\lim_{r\rightarrow\infty} f( r )=f_{\infty}$, where the constant $f_{\infty}$ is determined by the equation
\begin{equation}
1-f_{\infty}+\alpha f_{\infty}^2+(\mu+\xi) f_{\infty}^3=0\, .
\end{equation}
%Observe that there is no contribution from the combination $81\mathcal{S}_5+1106 \mathcal{Z}_5$, and indeed this is the reason to introduce it.
Then, the solution is asymptotically AdS with radius $\tilde L=L/\sqrt{f_{\infty}}$.  In this case, we fix  $N=1/\sqrt{f_{\infty}}$, which in the holographic context would be equivalent to imposing the speed of light to be $c=1$ in the dual CFT \cite{Buchel:2009sk}. Observe also that the effective gravitational constant reads  \cite{Aspects}
\begin{equation}
G_{\rm eff}=\frac{G}{1-2\alpha f_{\infty}-3(\mu+\xi) f_{\infty}^2}\, ,
\end{equation}
which has to be positive in order for the theory not to propagate ghosts. 
%Using the results of previous sections and \cite{Aspects}, it is straightforward to see that the matter-coupled linearized equations of \req{5dsg} read simply 
%\begin{equation}
%G_{ab}^L=8 \pi G_{\rm eff} T^L_{ab}\,,
%\end{equation}
%which are of course the same as for Einstein gravity up to the effective Newton constant.

Let us now consider a large $r$ expansion of $f(r)$ of the form
\begin{equation}
f( r )=f_{\infty}+\frac{G_{\rm eff}\omega^4}{G r^4}+f_1( r )\, ,
\end{equation}
where we have made explicit the leading terms and considered $f_1$ as a small correction. If we plug this in \req{5Dequation}, expand linearly in $f_1$ and keep only the leading terms when $r\rightarrow\infty$, we obtain
\begin{equation}
\frac{6\tilde{\xi}}{r^2}f_1''-\frac{G^2}{G_{\rm eff}^2\omega^4}f_1+\frac{G_{\rm eff} \omega^4}{G r^8}\left[\alpha+3f_{\infty}(\mu+\xi-20\tilde{\xi})\right]=0\,.
\end{equation}
This equation has the following approximate solutions when $r$ is large:
\begin{equation}
f_1( r )\sim A \exp\left[\frac{G r^2}{G_{\rm eff}\omega^22\sqrt{6\tilde{\xi}}}\right]+B\exp\left[-\frac{G r^2}{G_{\rm eff}\omega^22\sqrt{6\tilde{\xi}}}\right]+\frac{G^3_{\rm eff} \omega^8}{G^3 r^8}\left[\alpha+3f_{\infty}(\mu+\xi-20\tilde{\xi})\right]+\ldots,
\label{asymptoticf}
\end{equation}
for some integration constants $A$ and $B$.
Now, since we want $f_1\rightarrow 0$ as $r\rightarrow +\infty$, we must set $A=0$. This means that the boundary condition at infinity fixes one of the integration constants in \req{5Dequation}. Note however that this cannot be done if $\xi<0$ because in that case we obtain oscillating solutions which do not decay. Therefore, we must choose $\xi>0$. 

The other condition we need to impose is regularity of the solution at the horizon (where $f(r_h)=0$), which means that $f$ is infinitely differentiable there. Let us then Taylor expand $f(r)$ around the horizon $r=r_h$ as
\begin{equation}
f( r ) =\sum_{n=1}^{\infty}a_n (r-r_h)^n\, ,
\label{Taylorf}
\end{equation}
where $a_n=f^{(n)}(r_h)/n!$. Now, the parameter $a_1$ can be related to the temperature of the black hole. Indeed, by considering the near-horizon Euclidean metric we can see that the Euclidean time $\tau=i t$ must have a periodicity of $(4\pi L^2\sqrt{f_{\infty}})/(r_h^2a_1)$, so that we identify
\begin{equation}
a_1=4\pi T\frac{L^2\sqrt{f_{\infty}}}{r_h^2}\, .
\end{equation}
Now we plug \req{Taylorf} into \req{5Dequation} and we solve order by order in $(r-r_h)$. The first two terms give us the following constraints,
\begin{eqnarray}
1-32\tilde{\xi}\frac{(\pi T L^2\sqrt{f_{\infty}})^3}{r_h^3}-\frac{\omega^4}{r_h^4}&=&0\, ,\\
-\frac{\pi TL^2\sqrt{f_{\infty}}}{r_h}+36\tilde{\xi}\frac{(\pi TL^2\sqrt{f_{\infty}})^3}{r_h^3}+\frac{\omega^4}{r_h^4}&=&0\, ,
\end{eqnarray}
which determine $r_h$ and $T$ as functions of $\omega$. Remarkably, only the new cubic interaction $\mathcal{S}_5$ contributes to these equations,\footnote{This was also observed in \cite{Hennigar:2017ego} for the general $D$ asymptotically flat case.} which might have interesting holographic consequences. The rest of equations fix the coefficients $a_3$, $a_4$ and so on,   once $a_2$ is specified. Therefore, there is one single free parameter, which means that the regular horizon condition is already fixing one integration constant. Then, similarly to the $D=4$ asymptotically flat Einsteinian cubic gravity case studied in \cite{PabloPablo2,Hennigar:2016gkm}, the value of $a_2$ must be chosen so that the solution is asymptotically AdS, this is, so that the growing exponential mode in \req{asymptoticf} is not excited. This value can be found by performing a numerical analysis, in which we choose a value for $a_2$ and then solve numerically \req{5Dequation}. For an arbitrary choice of $a_2$, the solution rapidly diverges as $r\rightarrow\infty$, but for an appropriate value of this parameter the solution can be extended up to a sufficiently large $r$, in which the approximation \req{asymptoticf} holds.  Indeed, the analysis shows that there is only one value of $a_2$ for which $f( r )\rightarrow f_{\infty}$ when $r\rightarrow\infty$, and hence, there is a unique asymptotically AdS black hole solution.

From the previous expressions we find that the temperature and the horizon radius are 
\begin{equation}
\label{T5D}
T=\frac{\omega}{\pi L^2\sqrt{f_{\infty}}}\frac{\gamma(\tilde{\xi})}{(9-8\gamma(\tilde{\xi}))^{1/4}}\, , \quad  r_h=\frac{\omega}{(9-8\gamma(\tilde{\xi}))^{1/4}}\, ,
\end{equation}
where $\gamma(\tilde{\xi})$ is the solution to the cubic equation
\begin{equation}
1-\gamma(\tilde{\xi})+4\tilde{\xi} \gamma(\tilde{\xi})^3=0\, .
\end{equation}
We must demand $1\le \gamma(\tilde{\xi})< 9/8$ in order to get real and positive quantities, which imposes the constraint $0\le\tilde{\xi}<16/729$ (or, equivalently, $0\le \xi < 632/25515$).
Now we can compute the entropy using Wald's formula \cite{Wald:1993nt},
\begin{equation}
\mathsf{S}=-2\pi\int_{h}d^3x\sqrt{g_h}\frac{\delta \mathcal{L}}{\delta R_{abcd}}\epsilon_{ab}\epsilon_{cd}\, ,
\end{equation}
where the $\epsilon_{ab}$ are binormals to the horizon and $g_h$ is the determinant of the induced metric on that surface.
Applying this expression to our theory and evaluating on the horizon using \req{Taylorf}, we find
\begin{equation}
\mathsf{S}=\frac{A}{4 G_5}\left[1-36\pi^2\tilde{\xi} f_{\infty}\frac{L^4 T^2}{r_h^2}\right]\, . 
\end{equation}
The area of the horizon is $A=r_h^3 V_3/L^3$, where $V_3=\int d^3x$ is the infinite transverse volume. Observe that the Gauss-Bonnet and Quasitopological terms do not modify the usual Einstein gravity area-law at all \cite{Quasi}. Using the result for the temperature \req{T5D}, we find that the entropy and energy densities $s=\mathsf{S}/V_3$ and $\varepsilon$ can be written as
%Note again that the only contributions to $T$ and $\mathsf{S}$ coming from the Gauss-Bonnet and Quasitopological terms appear through $f_{\infty}$, while $\mathcal{S}_5$ introduces 
\begin{equation}
s(T)=\frac{(9-8 \gamma(\tilde{\xi}))(\pi L\sqrt{f_{\infty}})^3}{4 \gamma(\tilde{\xi})^4G_5}T^3\, ,\quad \varepsilon(T)=\frac{3(9-8\gamma(\tilde{\xi}))(\pi L\sqrt{f_{\infty}})^3}{16\gamma(\tilde{\xi})^4G_5}T^4\, ,
\end{equation}
where we used the first law $d\varepsilon= Tds$. From this, it is possible to obtain the free energy density
\begin{equation}
\mathcal{F}=\varepsilon - T s=- \frac{(9-8\gamma(\tilde{\xi}))(\pi L\sqrt{f_{\infty}})^3}{16\gamma(\tilde{\xi})^4G_5}T^4\, ,
\end{equation}
which can be alternatively obtained from the Euclidean on-shell action $\mathcal{F}=(S_{\rm E}-S_{\rm E}^0) T/V_3$, where $S_{\rm E}^0$ is the Euclidean version of \req{5dsg} evaluated on the AdS$_5$ background.

Observe that the above expressions, which contain an overall factor $(9-8\gamma(\tilde{\xi}))/\gamma(\tilde{\xi})^4$ with respect to their Gauss-Bonnet and Quasitopological gravity counterparts \cite{Buchel:2009sk,Quasi}, satisfy the relation
\begin{equation}
\varepsilon=\frac{3}{4}Ts\, ,
\end{equation}
as expected for a four-dimensional CFT at finite temperature.

\section{Final comments}\label{disc}
In this paper we have presented various results on the structure and properties of higher-derivative gravities admitting simple black hole solutions. A summary of our main results can be found in section \ref{main}, so we shall not repeat ourselves here. Let us nevertheless make some final comments regarding possible future work.

First, it would be interesting to find  a general proof of Conjecture \ref{conj} which, we think, is extremely likely to be correct in view of the strong evidence accumulated so far for various theories  \cite{Wheeler:1985nh,Wheeler:1985qd,Boulware:1985wk,Cai:2001dz,Dehghani:2009zzb,deBoer:2009gx,Quasi,Quasi2,Dehghani:2011vu,PabloPablo,Hennigar:2016gkm,PabloPablo2,Cisterna:2017umf,Hennigar:2017ego}, and further supported by the results presented in sections \ref{bho1} and \ref{planar}. It is also likely that a proof for Conjecture \ref{conji2} can be found, although the physical interest of such result would be probably smaller.

In addition, one could try to extend the discussion of section \ref{bho1} to time-dependent situations, which could lead to elucidate whether these theories satisfy a version of Birkhoff's theorem, known to hold for some higher-derivative gravities \cite{Oliva:2011xu,Oliva:2012zs}.

Let us also point out that using the machinery developed here, it should be straightforward to construct the quartic version (and higher-order versions) of Generalized quasitopological gravity \cite{Hennigar:2017ego}. Preliminary results suggest that this theory will contain, apart from all the terms in \req{gQuasi2}, the quartic quasitopological term \cite{Dehghani:2011vu} plus a single extra new term (defined up to quartic terms which do not contribute to the equations of motion for the general ansatz \req{Nf}). Naturally, a more ambitious goal would be a full characterization of all $\mathcal{L}(g^{ab},R_{abcd},\nabla_{e}R_{abcd},\ldots)$ theories satisfying the hypothesis of Result \ref{theo} and their black hole solutions.

Along different lines, it would be interesting to perform holographic studies for some of the recently constructed theories satisfying the hypothesis or Result \ref{theo}, such as four-dimensional Einsteinian cubic gravity \cite{PabloPablo} or Generalized quasitopological gravity \cite{Hennigar:2017ego}. In view of their special properties, these theories seem to be ideal candidates for such applications.

\section*{Acknowledgments} 

\noindent We are happy to thank Adam Bzowski, Rob Myers, Julio Oliva, Tom\'as Ort\'in, Pedro F. Ram\'irez and Manus Visser for useful comments. The work of PB was supported by a postdoctoral fellowship from the Fund for Scientific Research - Flanders (FWO). The work of PAC was supported by a ``la Caixa-Severo Ochoa'' International pre-doctoral grant and in part by the Spanish Ministry of Science and Education grants FPA2012-35043-C02-01 and FPA2015-66793-P and the Centro de Excelencia Severo Ochoa
Program grant SEV-2012-0249.

\renewcommand{\leftmark}{\MakeUppercase{Bibliography}}
\phantomsection
\bibliographystyle{JHEP}
\bibliography{Gravities}
\label{biblio}

\end{document}